\documentclass[10pt, paper=a4, UKenglish]{article}
\usepackage{graphicx}
\graphicspath{{fig/}}
\usepackage{bm}
\usepackage[sicmds]{hepunits}
\usepackage{amssymb}
\usepackage{amsmath}
\DeclareSIUnit{\clight}{\text{\ensuremath{c\hspace{0.2ex}}}}
\DeclareSIUnit{\eVc}{\eV\hspace*{-0.25ex}\per\clight}
\DeclareSIUnit{\eVcsq}{\eV\hspace*{-0.25ex}\per\clight^2}
\DeclareMathSymbol{\shortminus}{\mathbin}{AMSa}{"39}

\DeclareSIUnit\barn{b}

\newcommand{\smallsim}{\smallsym{\mathrel}{\sim}}

\makeatletter
\newcommand{\smallsym}[2]{#1{\mathpalette\make@small@sym{#2}}}
\newcommand{\make@small@sym}[2]{%
  \vcenter{\hbox{$\m@th\downgrade@style#1#2$}}%
}
\newcommand{\downgrade@style}[1]{%
  \ifx#1\displaystyle\scriptstyle\else
    \ifx#1\textstyle\scriptstyle\else
      \scriptscriptstyle
  \fi\fi
}
\makeatother

\def\Title#1{\begin{center} {\Large #1 } \end{center}}
\def\Author#1{\begin{center}{ \sc #1} \end{center}}
\def\Address#1{\begin{center}{ \it #1} \end{center}}

\newcommand\pubblock{\rightline{\begin{tabular}{l} Proceedings of the CTD 2023\\ \pubnumber\\
         \pubdate  \end{tabular}}}

\newenvironment{Abstract}{\begin{quotation} \begin{center} 
             \large ABSTRACT \end{center}\bigskip 
      \begin{center}\begin{large}}{\end{large}\end{center} \end{quotation}}

\newenvironment{Presented}{\begin{quotation} \begin{center} 
             PRESENTED AT\end{center}\bigskip 
      \begin{center}\begin{large}}{\end{large}\end{center} \end{quotation}}

\def\Acknowledgements{\bigskip  \bigskip \begin{center} \begin{large}
      \bf ACKNOWLEDGEMENTS \end{large}\end{center}}

\def\beq{\begin{equation}}
\def\eeq#1{\label{#1}\end{equation}}
\def\eeqn{\end{equation}}

\def\beqa{\begin{eqnarray}}
\def\eeqa#1{\label{#1}\end{eqnarray}}
\def\eeqan{\end{eqnarray}}

\let\bar=\overbar

\def\ie{{\it i.e.}}

\def\Dslash{\not{\hbox{\kern-4pt $D$}}}
\def\dslash{\not{\hbox{\kern-2pt $\del$}}}

\def\msb{{\bar{\ssstyle M \kern -1pt S}}}

\def\s#1{\widetilde{#1}}

\usepackage{color}
\usepackage{lineno}
\usepackage{subfig}
\usepackage{hyperref}

\newcommand\pubnumber{PROC-CTD2023-16}

\newcommand\pubdate{April, 8 2024}

\def\affiliation{
on behalf of the LHCb RTA project\\~\\
$ ^1$ INFN Sezione di Pisa, Pisa, Italy \\
$ ^2$ Universit\`a degli Studi di Siena, Siena, Italy \\
$ ^3$ INFN Sezione di Ferrara, Ferrara, Italy\\
$ ^4$ European Organization for Nuclear Research (CERN), Geneva, Switzerland\\
$ ^5$ Scuola Normale Superiore, Pisa, Italy \\
$ ^6$ INFN Sezione di Cagliari, Monserrato, Italy\\
$ ^7$ Universit\`a di Ferrara, Ferrara, Italy\\
$ ^8$ Universit\`a di Pisa, Pisa, Italy\\
$ ^9$ Physikalisches Institut, Ruprecht-Karls-Universitat Heidelberg, Heidelberg, Germany. Formerly with University of Chinese
Academy of Sciences, Beijing, P.R.China}

\newcommand{\conference}{Connecting the Dots Workshop (CTD 2023)\\
October 10-13, 2023}

\usepackage{fancyhdr}
\definecolor{mygrey}{RGB}{105,105,105}
\begin{document}


\large
\begin{titlepage}
\pubblock

\vfill
\Title{A real-time demonstrator of track reconstruction with FPGAs at LHCb}
\vfill

\Author{Francesco~Terzuoli$^{1,2}$, Wander~Baldini$^{3,4}$, Giovanni~Bassi$^{1,5}$, Andrea~Contu$^6$, Riccardo~Fantechi$^1$, Sofia~Kotriakhova$^{3,7}$, Federico~Lazzari$^{1,8}$, Michael~Joseph~Morello$^{1,5}$, Giovanni~Punzi$^{1,8}$, Giulia~Tuci$^9$}
\Address{\affiliation}
\vfill

\begin{Abstract}
The upgraded LHCb detector has started its Run 3 of data taking in 2022, with a completely overhauled DAQ system, reading out and processing the full detector data at every LHC bunch crossing (\SI{30}{\MHz} average rate). At the same time, an intense R\&D activity is taking place, with the aim of further improving the real-time data processing performance of LHCb, in view of “Upgrade II”, where luminosity will be increased.
In this work, we describe the experience gained with a prototype device for a \SI{30}{\MHz} real-time tracking in the LHCb VELO detector, implemented in state-of-art PCIe-hosted FPGA cards interconnected by fast optical links.
The system has been processing live LHCb data opportunistically during physics data taking, thanks to a dedicated TestBed facility fed by the experiment monitoring system. We describe, amongst other things, the system used to organise and optimise the high-speed distribution of data to the components, and the synchronisation with the most updated alignment constants to be used in track reconstruction.
\end{Abstract}

\vfill

\begin{Presented}
\conference
\end{Presented}
\vfill
\end{titlepage}
\def\thefootnote{\fnsymbol{footnote}}
\setcounter{footnote}{0}
%

\normalsize 


\section{Introduction}
\label{intro}
Heterogeneous computing architectures are becoming more and more sought after by HEP experiments, in view of ever-increasing data-flow and complexity, while at the same time Moore's law is showing signs of slowing down~\cite{Moore7878935}.
LHCb resorts to this kind of cutting-edge solutions in the first stage of the LHCb High Level Trigger (HLT1) for Run~3, which is GPU-based~\cite{LHCbCollaboration:2717938}. With the approaching of Upgrade II, where the luminosity will be increased by a further factor of 5 to 10 with respect to Run~3~\cite{lhcbupgrade2}, a Coprocessor TestBed has been established, in order to test new processing solutions in realistic DAQ conditions during the Run~3, with live LHCb data being opportunistically delivered to this facility.

One such solution under development is a highly-parallelised custom tracking processor based on the ``Artificial Retina'' architecture.
The ``Artificial Retina'' architecture~\cite{RISTORI2000425} takes advantage of FPGA parallel computational capabilities, by distributing the processing of each event over an array of FPGA cards, interconnected by an optical network. This is expected to allow operation in real-time at the full LHC collision rate, with no need for time-multiplexing or extra buffering. 
Several lab prototypes have already been built and tested during the past years of R\&D, allowing to separately test the various functionalities of the system on a reduced scale~\cite{lazzari_twepp2021}. In this work, we report the results obtained with a more complete and realistic demonstrator of this technology, covering a significant chunk of an existing LHCb detector, and we describe the system used to organise and optimise the high-speed
distribution of data to the components. How to synchronise the tracks reconstruction with the most updated alignment constants and operational experience on live LHCb data is also discussed.

\section{The ``Artificial Retina''}
\label{sec:artificial_retina}
The ``Artificial Retina'' architecture\footnote{Henceforth, it will be referred simply as Retina, for sake of brevity and readability.} is an arrangement of parallel computing units (cells) fed by a custom distribution network, which are programmed to perform, at a very high rate, a computation resembling the ``Hough transform'' \cite{Hough1}, a mathematical approach for finding lines in image processing tasks.

\begin{figure}[!htb]
  \centering
  \subfloat[]{\includegraphics[width=0.27\linewidth]{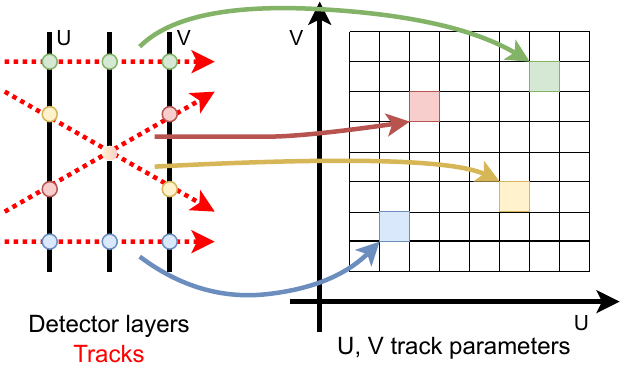}\label{subfig:step1}}
  \qquad
  \subfloat[]{\includegraphics[width=0.30\linewidth,trim={0 0 0 1.4cm},clip]{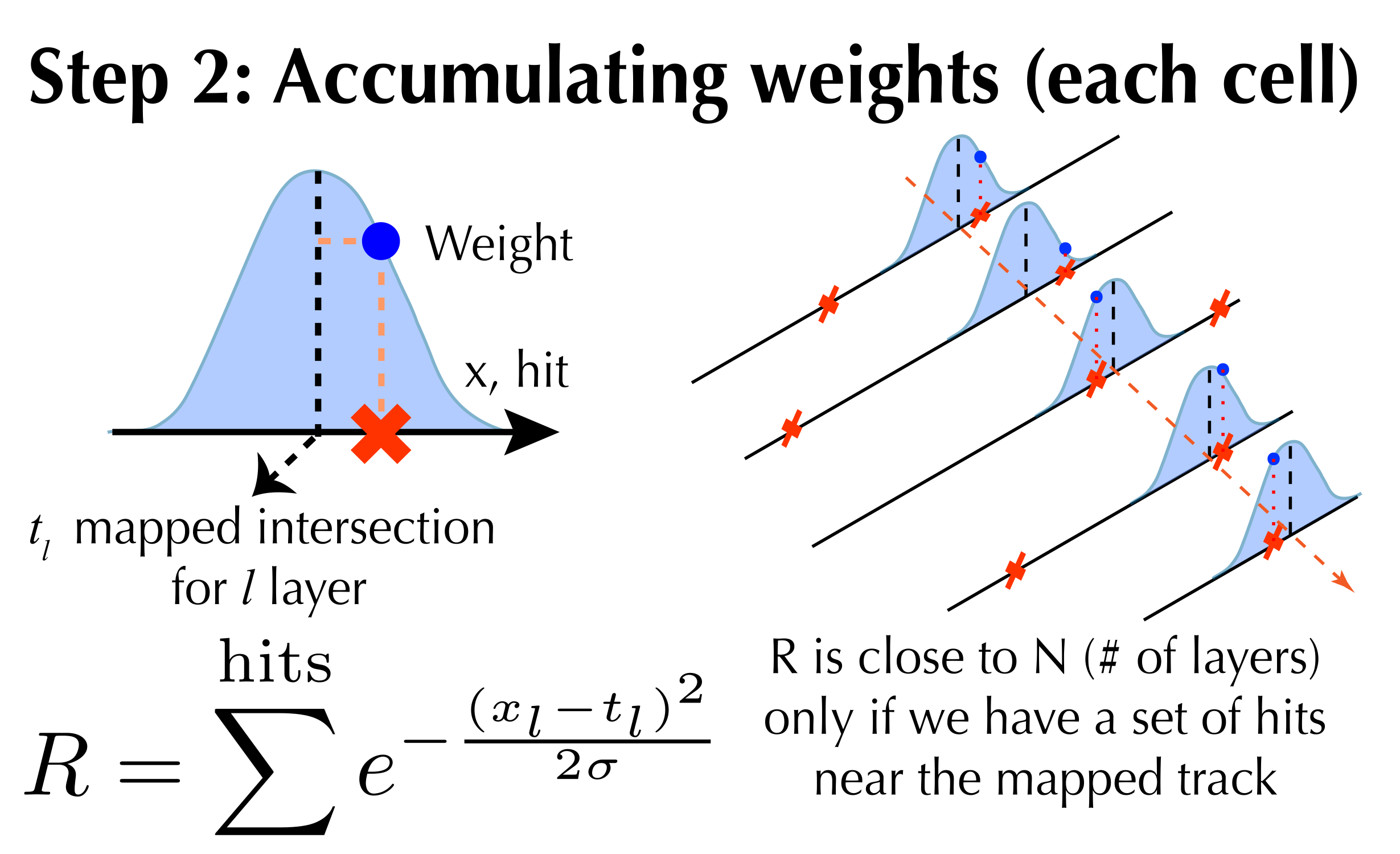}\label{subfig:step2}}
  \qquad
  \subfloat[]{\includegraphics[width=0.30\linewidth,trim={0 0 0 1.5cm},clip]{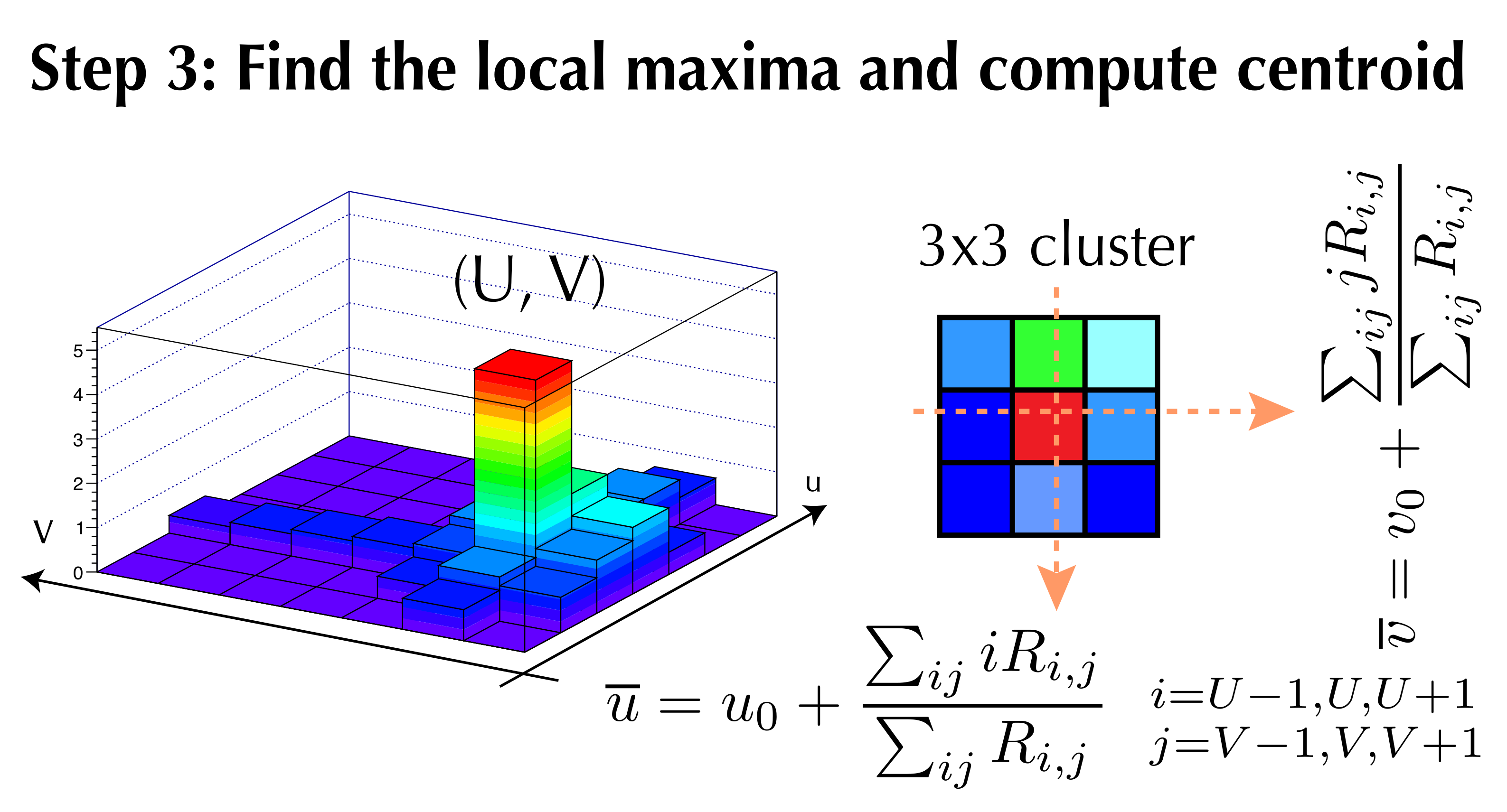}\label{subfig:step3}}
  \caption{The steps of the ``Artificial Retina" reconstruction algorithm: discretisation of the parameter space and creation of receptors (a); computation of cell ``excitation level" (b); local maxima finding, interpolation and parameter estimation of candidate track (c).}
  \label{fig:retina_steps}
\end{figure}

As first step, the track parameters space gets discretised so that it can be represented as a \emph{matrix of cells} $M\times N$. These parameters, indicated by $(u, v)$ in the following, are defined case-by-case and uniquely identify a track. The choice of a two-dimensional representation of the parameter space is naturally suggested by the two-dimensional resources arrangement inside FPGA chips. Then, $M\times N$ reference tracks are generated with their parameters set to the centre of each cell and their intersects with each layer of the chosen detector are computed (Figure~\ref{subfig:step1}); these intersects are called \emph{receptors}.

Each cell hosts a computational \emph{engine}, entirely independent from the others and implemented in a modest amount of logic circuitry inside the FPGA. The engine checks whether the input hits from particles interacting with the detector are compatible with the reference tracks and returns a quantitative response indicative of such agreement.
To this purpose, each engine implements an accumulator, storing the ``excitation level" of the cell, and every hit contribute to it with a weight dependent of the distance between the hit itself and the receptor on the same layer (Figure~\ref{subfig:step2}). This ``excitation function" can take the form of a Gaussian distribution with mean set to the receptor coordinates and variance being a tuning parameter related to the dimension of the cell \cite{chep_giulia}. Another tunable parameter, denominated ``search distance", is the maximum distance between hit and receptor beyond which the hit carries no contribution to the accumulator by setting its weight to zero.

After all hits from an event are received, the Track Processing Units (TPUs), which are sub-matrices of adjacent cells implemented on the same FPGA board, search for local maxima and interpolate the response of adjacent cells, so that the position of their centroid is taken as an estimation of the candidate track parameters (Figure~\ref{subfig:step3}).

\subsection{The distribution network}
\label{subsec:distribution_network}
Distributing hits from DAQ readout lines to the relevant cells constitutes a core task within Retina and due to this the distribution network occupies a central role. Each hit must be routed to engines where it can contribute to the ``excitation level" with a non-null weight, but it can happen for one hit to be compatible with different tracks, and therefore needed to be routed to multiple cells: this results in the distribution network creating copies of the same hit. Handling and optimisation of this \emph{hit multiplication factor} is addressed in detail in Section~\ref{subsec:opti_switch}.

\begin{figure}[!htb]
  \centering
  \subfloat[]{\includegraphics[height=4cm]{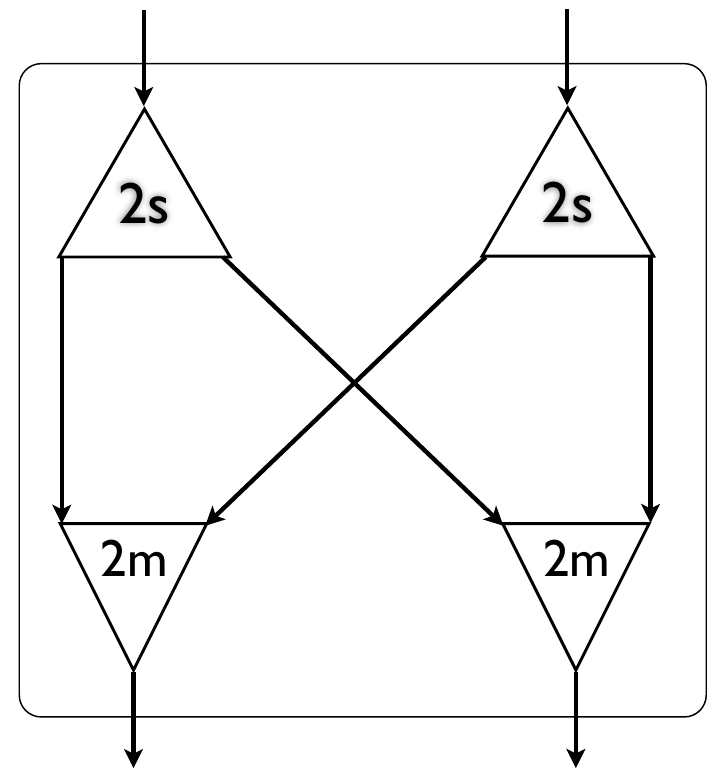}\label{subfig:disp_2d}}
  \qquad\qquad
  \subfloat[]{\includegraphics[height=4cm]{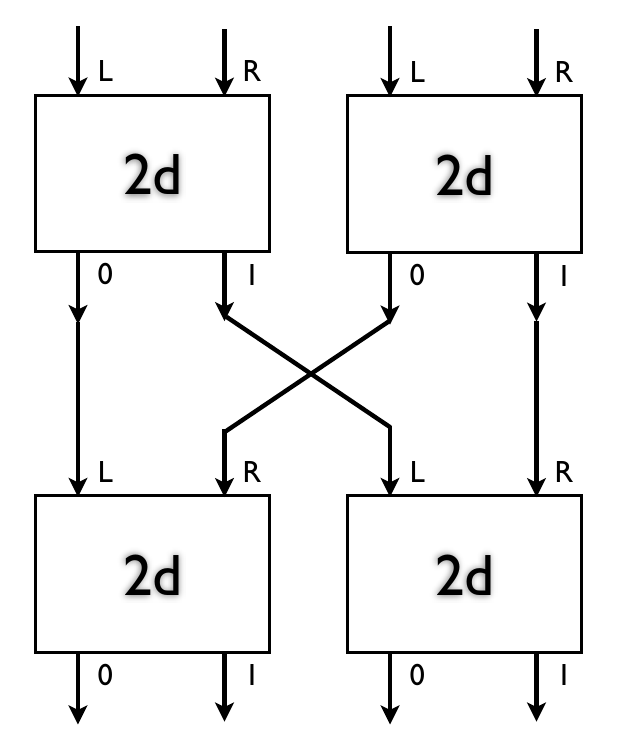}\label{subfig:disp_4d}}
  \qquad\qquad
  \subfloat[]{\includegraphics[height=4cm]{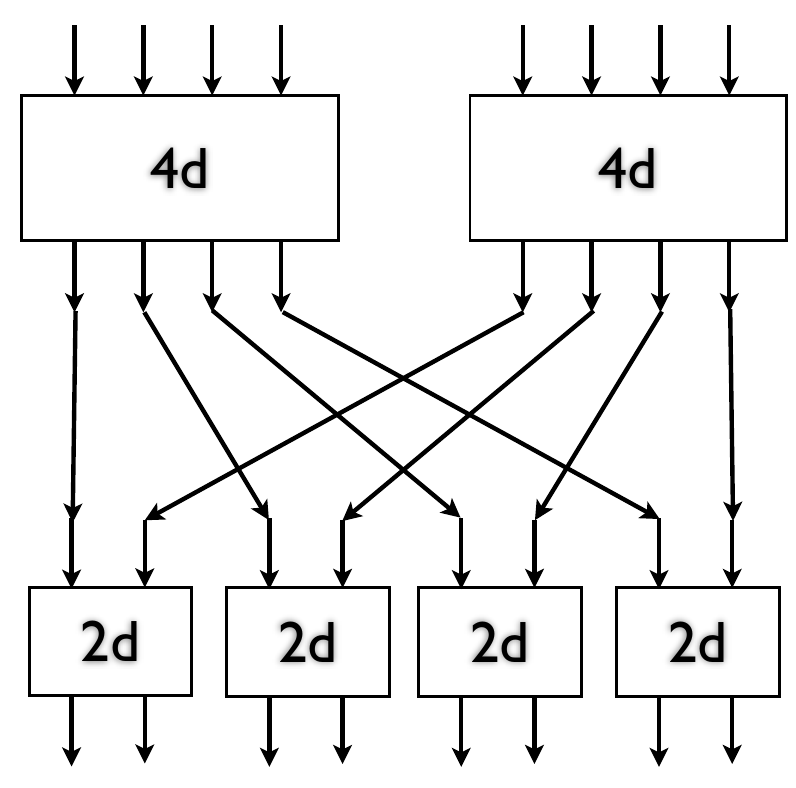}\label{subfig:disp_8d}}
  \caption{Architectures of a two-way dispatcher $2d$ (a), four-way dispatcher $4d$ (b) and 8-way dispatcher $8d$ (c), with mergers ($2m$) and splitters ($2s$) as elementary blocks.}
  \label{fig:switch_components}
\end{figure}

The distribution network is built according to a modular design. The basic blocks are a \emph{two-way splitter} ($2s$) and a \emph{two-way merger} ($2m$)~\cite{lazzari_thesis}. The former has one input line and two outputs: according to a Look-Up Table (LUT) incoming data can be routed to a specific output line, possibly both. The latter has two input lines and a single output, thus merging the incoming data flow into one line. By combining two splitters and two mergers interconnected as in Figure~\ref{subfig:disp_2d}, a \emph{two-way dispatcher} ($2d$) is obtained. Then, by arranging mergers, splitters and two-way dispatchers in a number of levels, a switch with an arbitrary number of input and output lines can be built as shown in Figures~\ref{fig:switch_components}(b-c) depicting the implementation of a four-way dispatcher ($4d$) and an eight-way dispatcher ($8d$).

With the entire system being spread over multiple boards and the whole network also being too wide to be implemented on a single chip, vertical segmentation of the network and lateral connections crossing segment boundaries are needed. Due to the complexity of this, the network is tripartite in sections and lateral connections are collapsed into two separate layers located at the crossing between the three different sections of the switch (Pre-Switch, Mid-Switch and Post-Switch). This allows for the sections to be implemented as arrays of separate entities, with a bundle of fast optical links for the lateral connections~\cite{lazzari_twepp2021}.

The result is a system that is arranged on an array of boards, identical in structure, that communicate with the host server for data exchange via PCIe. Point-to-point connections among different boards are accomplished by multi-fibre cables, which are spliced and routed to the desired destinations via a patch panel.

\section{Demonstrator}
\label{sec:demo}
A relevant test for proving the applicability of such system to real-life applications is building a sizeable demonstrator with the complete functionality expected from a future device, operating in realistic DAQ conditions. The target is demonstrating the capability of reconstructing a portion of an actual detector with a prototype working in real-time at nominal LHCb luminosity (\SI{2e33}{\invcmsqpersec}) for Run 3 and Run 4 \cite{TDRVelo} and integrated in the LHCb DAQ system.

The target detector chosen for this demonstrator is the VErtex LOcator (VELO) pixel detector. The main purpose of this detector is to reconstruct primary vertices and displaced secondary vertices, the latter being essential in $b\shortminus$tagging algorithms and, more in general, heavy flavour decays identification. The VELO detector is composed of 26 stations, 19 of which are placed in the forward region; each of them are made of two modules, one on the left-hand side of the detector, the other on the right-hand side, with the beampipe running between them \cite{TDRVelo}.

Several motivations contributed to this specific choice of detector and can be summarised as follows. The VELO is a complex, high resolution detector that is crucial to the experiment, coming
first in the reconstruction sequence; detailed simulations of Retina reconstruction of the VELO were already available at the time of starting the demonstrator project~\cite{chep_giulia}; its data is read out over a comparatively smaller number of lines in comparison to other LHCb detectors, allowing us to build a demonstrator covering a meaningful portion of the target detector with an affordable quantity of hardware; hit coordinate reconstruction is already implemented in VELO readout with a Retina-like approach, so that data is already partially prepared for being processed in the demonstrator \cite{Bassi:2023jpv}.

\begin{figure}[!htb]
  \centering
  \subfloat[]{\includegraphics[height=4.5cm]{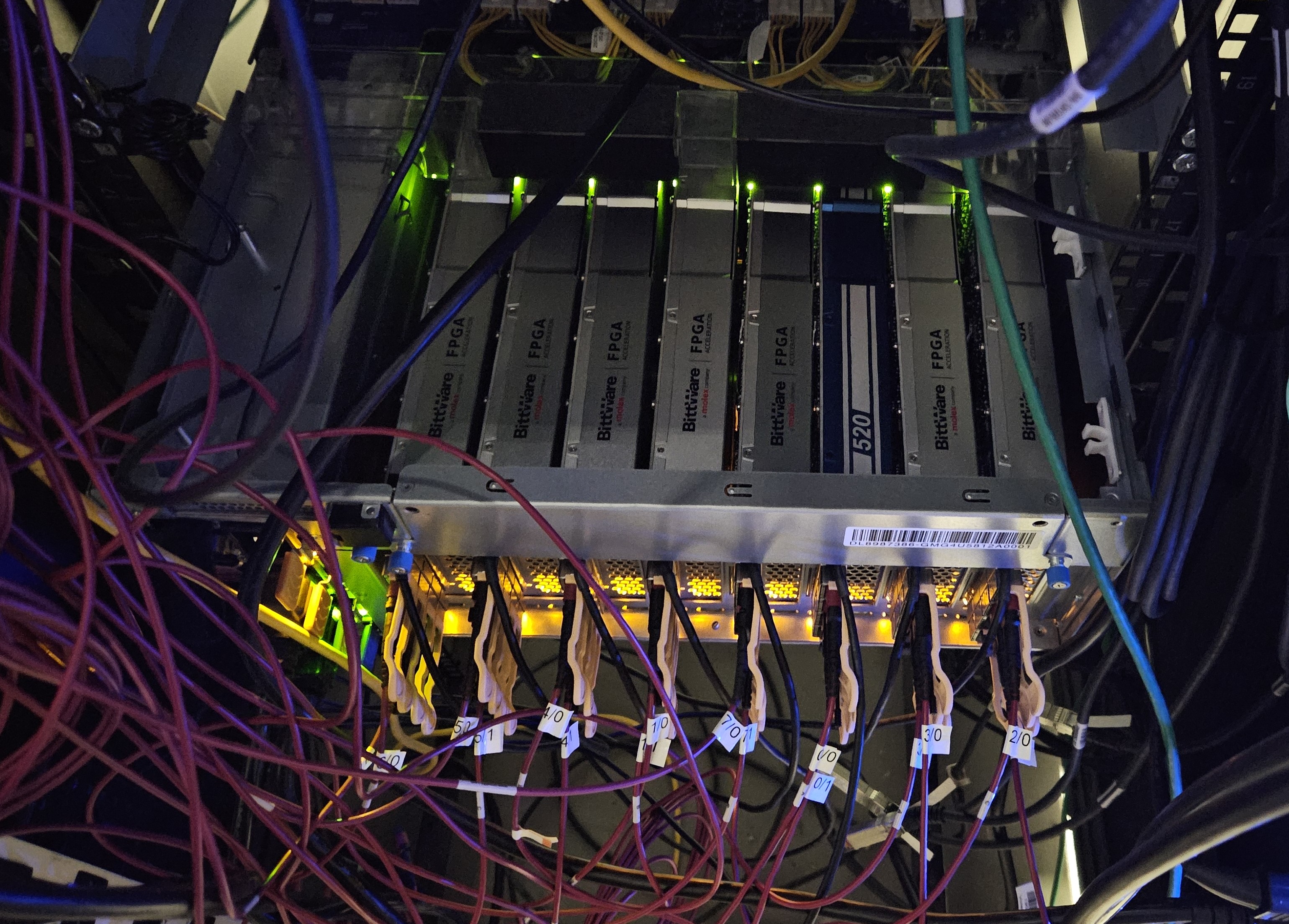}\label{subfig:demo}}
  \qquad\qquad
  \subfloat[]{\includegraphics[height=4.5cm]{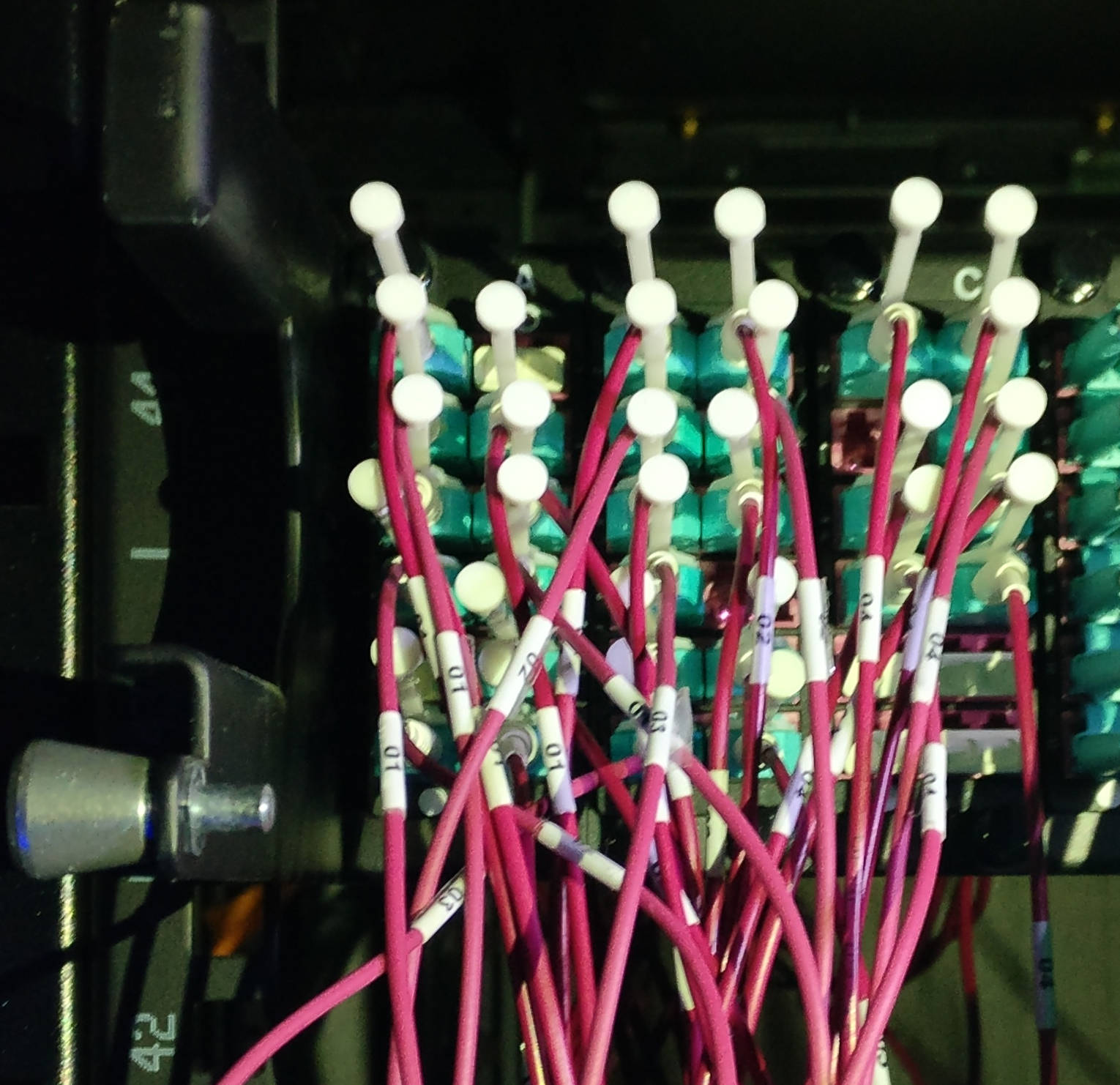}\label{subfig:8_node}}
  \caption{Photo (from above) of the Retina demonstrator fitted into a single server, located at the LHCb Coprocessor TestBed facility (a). Implementation of a 8-node full mesh optical network (b).}
\end{figure}

The prototype, located at the LHCb Coprocessor TestBed facility, comprises eight PCIe boards, each carrying a Stratix 10 FPGA, produced by Bittware/Molex and commercialised
under the name 520N\footnote{https://www.bittware.com/products/520n/}. Each one has a PCIe Gen3 x16 connector and 16 XCVRs with a maximum bandwidth of \SI{26}{\giga\bit/\s} each. The whole system fits within a single server (Figure~\ref{subfig:demo}), chosen of the same make and model of those used by the LHCb Event Builder to ease the integration with the LHCb DAQ environment.

The system currently covers 16 of the 38 VELO modules in the downstream region, all from the right-hand side of the detector, and maps a quadrant of the track parameters space.

\subsection{Implementation and optimisation of the distribution network}
\label{subsec:opti_switch}
The demonstrator physically implements $1/4$ of the system that would be needed for reconstructing the entire detector. This implies that the network distribution is devoid of the Pre-Switch sector, mentioned in Section \ref{subsec:distribution_network}. This stage of the switch and the subsequent layer of lateral connections handle inter-communication between different quadrants of the track parameters space, and therefore they are not needed in the present implementation. What remains is the Mid-Switch and Post-Switch vertically segmented among the eight FPGA boards, with a single layer of lateral connections between the different boards arranged in an $8\times 8$ full mesh network (Figure~\ref{subfig:8_node}). Each segment of the Mid-Switch is implemented as two side-by-side 4-way dispatcher, while each segment of the Post-Switch is an 8-way dispatcher.
This configuration leads to a total of 64 lines output from the switch, each delivering hits to a separate TPU. 

\begin{figure}[!htb]
  \centering
  \subfloat[]{\raisebox{-0.5cm}{\includegraphics[width=0.63\textwidth,height=7.8cm,clip]{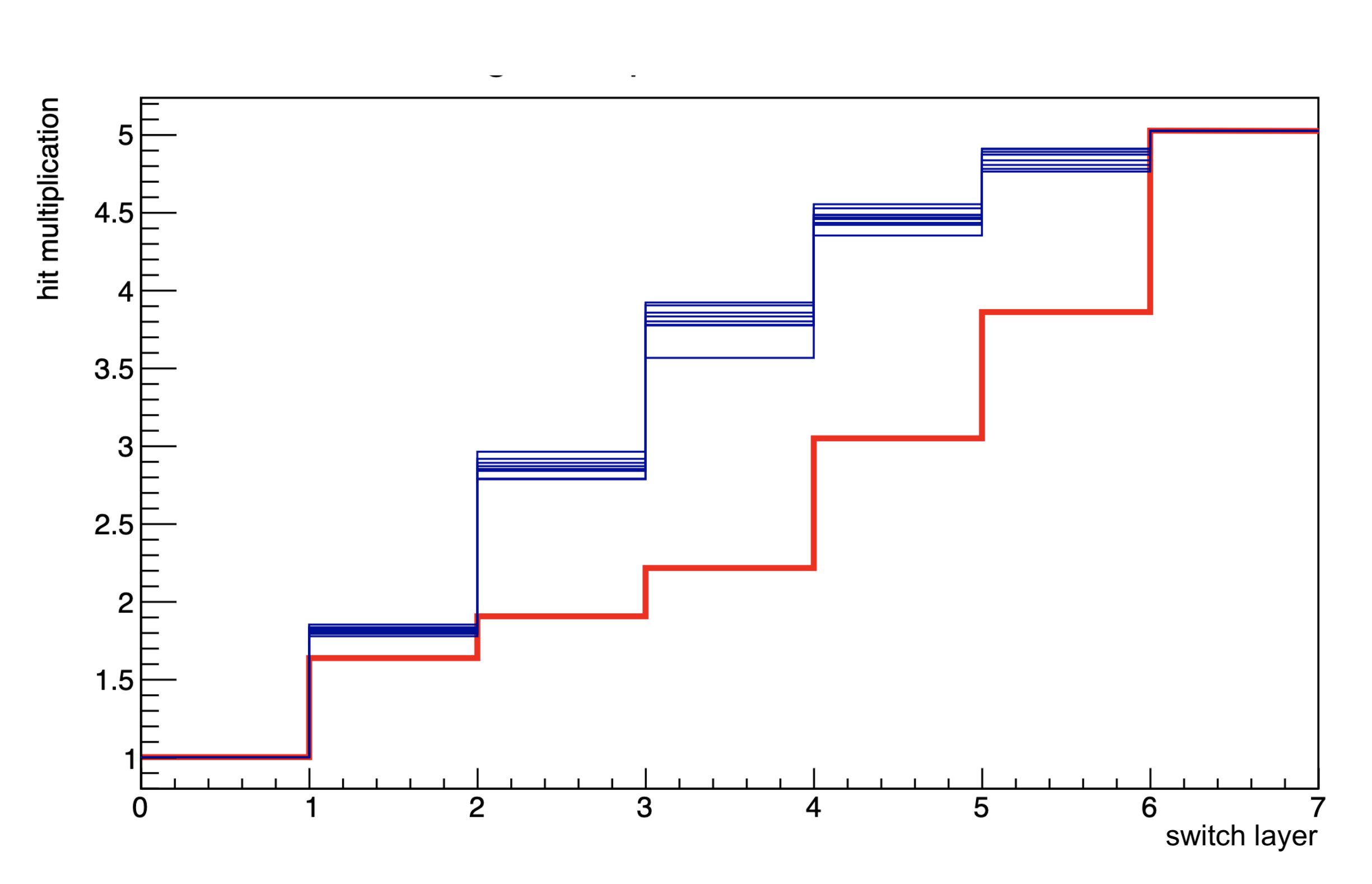}\label{subfig:hit_multi}}}
  \subfloat[]{\includegraphics[height=6.5cm]{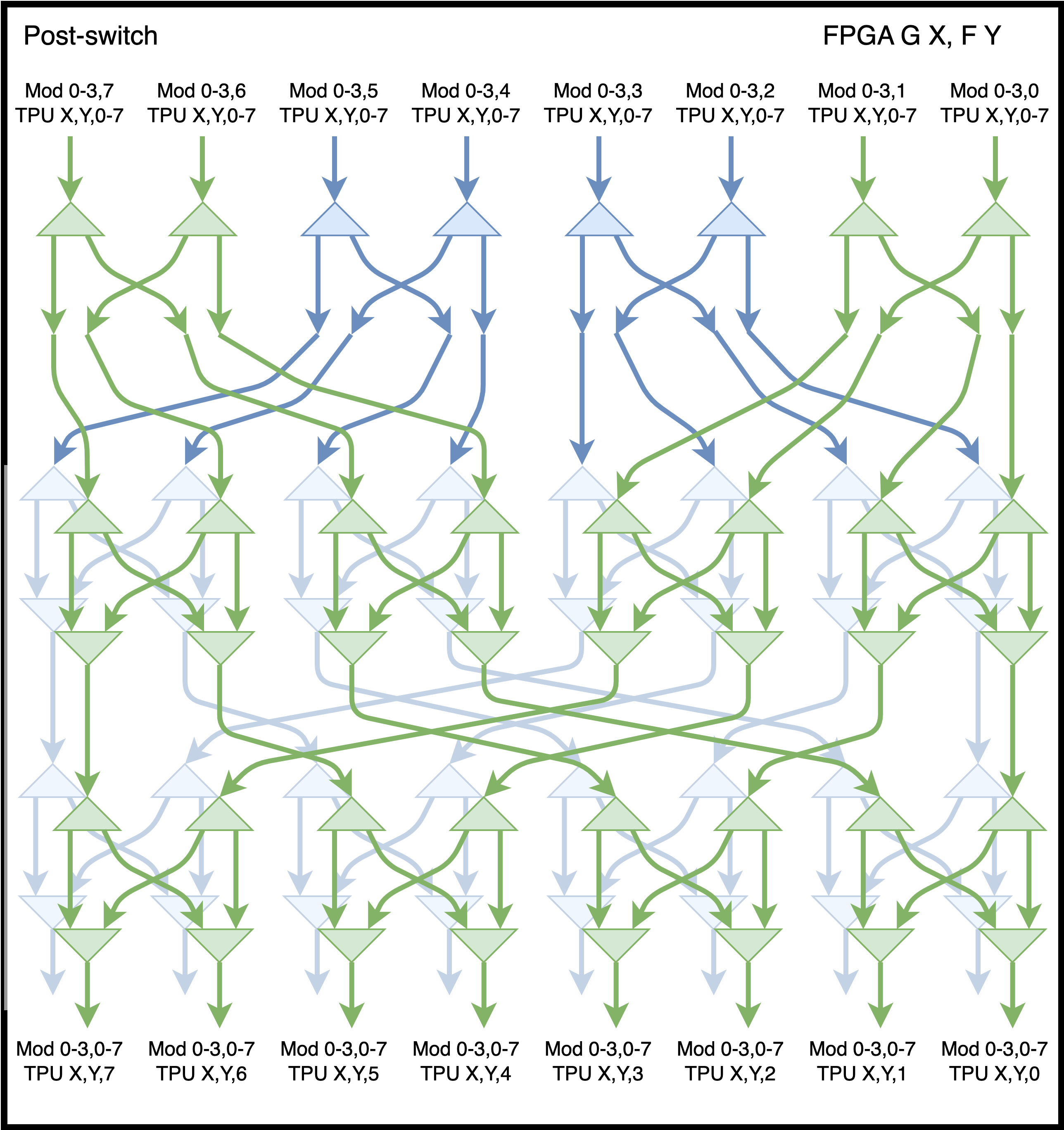}\label{subfig:switch16}}
  \caption{Hit multiplication factor at each layer of the switch: in blue random ordering of the TPUs, in red the optimised ordering (a); the layers are in increasing order from Pre-Switch ($0\shortminus2$) to Mid-Switch ($3\shortminus4$), and then Post-Switch ($5\shortminus7$). The Post-Switch as  $8\times 16$ dispatcher, employed in the Quad-Input design (b).}
  \label{fig:switch}
\end{figure}

The process of hit multiplication required by the Retina protocol significantly increases data volume, and can lead to congestion of the distribution network, hindering throughput performance. It is therefore important to optimise its programming. The switch delivers hits to a specific TPU only if at least one of its cell has a non-null weight. While the multiplication factor at the output of the switch is fixed by the size of the TPUs and the parameters of the cells (\ie\ ``search distance" and $\sigma$ of the Gaussian weight function), in the internal layers it can be tweaked by choosing a non-random order of the TPUs at the output of the switch. By iteratively pairing the TPUs with the most hits in common, and then recursively grouping such pairs (groups) of TPUs in the higher level(s),  hits multiplication can be delayed inside the switch to the layers nearest to the TPUs, where it has the least impact on performance (Figure~\ref{subfig:hit_multi}) \cite{thesis_novissimo}.

In addition, to limit line occupancy, each 8-way dispatcher of the Post-Switch has been replaced by a $8\times 16$ dispatcher by substituting the first layer of two-way dispatchers with a layer of splitters, and doubling the dispatchers in the lower layers (Figure~\ref{subfig:switch16}). Moreover, every FPGA board implements two instances of the same switch segment, each one handling half of the incoming hits; due to this, TPUs have four input lines.

With a coverage of 16 VELO modules each Mid-Switch segment is fed with data coming from two modules.

\subsection{Tests with simulated LHCb data at nominal Run 3 luminosity}
\label{subsec:MC}
Events generated by the official LHCb simulation at Run~3 conditions, with centre of mass energy $\sqrt{s}=\SI{14}{\TeV}$  and instantaneous luminosity $\mathcal{L} = \SI{2e33}{\invcmsqpersec}$, are loaded into internal RAMs of the demonstrator and they are read in-loop to provide a continuous flux of data. The output of the demonstrator is sent back to the host server via PCIe. Exact adherence to the results expected from custom developed C++ bitwise emulation is tested and event separation is constantly checked.

The system ran uninterruptedly for 10 days without errors, before stopping due to the corruption of the word separating two consecutive events, occurred in the communication between the boards. During this time no discrepancies were detected in the reconstructed tracks w.r.t. the output of the software emulation.

The demonstrator was able to reach an event rate of \SI{19.0}{\MHz}. As of reference the maximum throughput with a non-optimised switch that sends all hits to all TPUs was \SI{170}{\kHz}, at the full nominal luminosity of Run 3.

This is an unprecedented throughput, and while it still falls short of the LHC average beam crossing rate of \SI{30}{\MHz} by a factor of $\times 1.58$, the target event rate (\SI{30}{\MHz}) is expected to be within reach of the current hardware with some optimisation both of firmware and hits distribution architecture (\ie\ hit duplication and line occupancy balancing in the switch, final output logic chain, and clock frequency). The combined effect of this optimisation is estimated to produce a factor $\times 1.6$ in throughput.

\subsection{Processing live LHCb data from \texorpdfstring{$\bm{pp\shortminus}$}{pp-}collisions}

\begin{figure}[!htb]
\begin{center}
\includegraphics[width=0.9\textwidth]{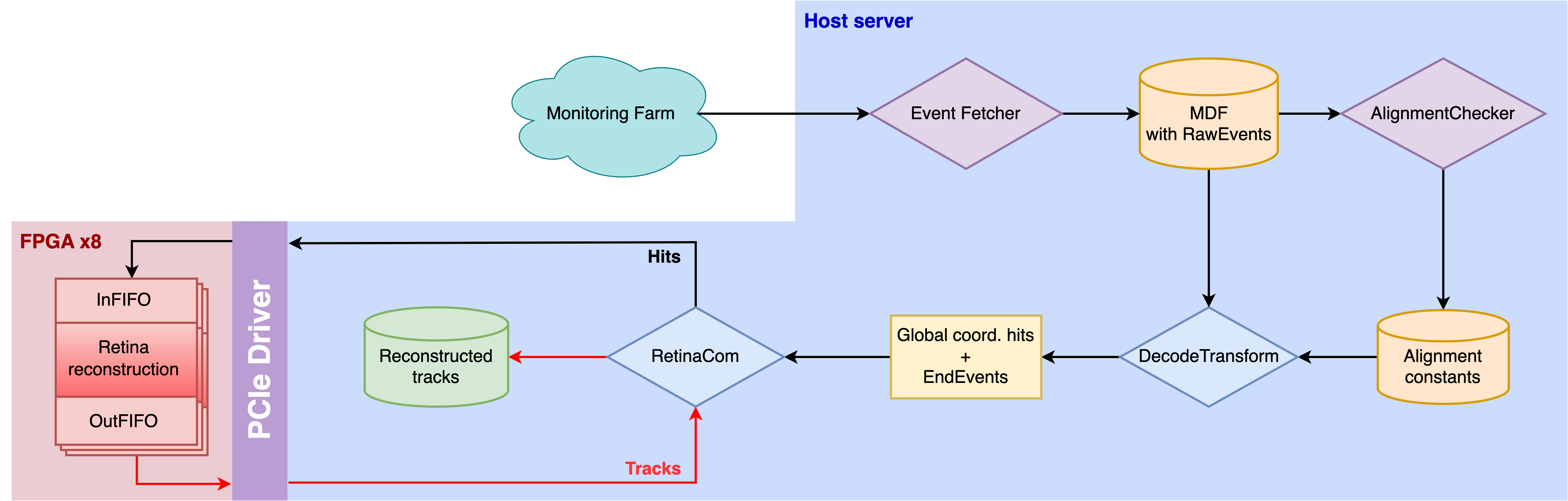}
\caption{Dataflow from the monitoring farm to the demonstrator in the live data test.}
\label{fig:live_data}
\end{center} 
\end{figure}

The final test of the demonstrator was about its capability of processing real data produced by LHC collisions happening at IP8, in real time.
Having proved the ability of the system to run reliably for days at full speed by feeding it with the same simulated sample, we wanted to test its behaviour when dealing with ever-changing real-collision data, even if the rate of such events available at the TestBed facility is currently limited by the low bandwidth of the connection (\SI{1}{\giga\bit/\s}).
In order to do so, a dedicated data-chain has been developed (Figure \ref{fig:live_data}): RawEvents arriving from the monitoring farm at the rate of \SI{1}{\kHz} are buffered and stored on disk in files of approximately \SI{2}{GB} each (corresponding to $\smallsim 2$ minutes of data taking). The ``AlignmentChecker" script connects to the LHCb Condition Database and delivers the real-time detector alignment constants, computed by the LHCb framework~\cite{borghi_align}, to ``DecodeTransform" which extracts VELO RawBanks from the file, decodes them and transforms the hits from local coordinate system of the sensors to the global coordinate system of LHCb. Then, these hits are sent in eight different streams (one per FPGA) to ``RetinaCom", which handles the communication between the host server and the FPGAs.
As in previous tests, each FPGA receives data from two VELO modules; however, instead of loading hits to the internal FPGAs' RAMs, they are directly loaded into the internal input FIFOs.
Such communication is handled by the PCIe stock driver provided by the boards' manufacturer.

Hit loading is carried out without any constraints on event synchronisation.
Subsequently, the demonstrator processes the hits exactly as in the simulated data test, monitoring events separation.
Then the output FIFOs are read and the reconstructed tracks stored onto disk.
Access to the internal FIFOs is carried out in multi-thread mode (one thread per FPGA) in order to speed up the process.
The Run Number and the original name of the incoming file are persisted throughout the whole chain, to allow, in subsequent offline studies, a direct comparison between the ``Retina'' performance on real collision data and standard LHCb reconstruction.

\begin{figure}[!htb]
  \centering
  \subfloat[]{\includegraphics[height=6.8cm]{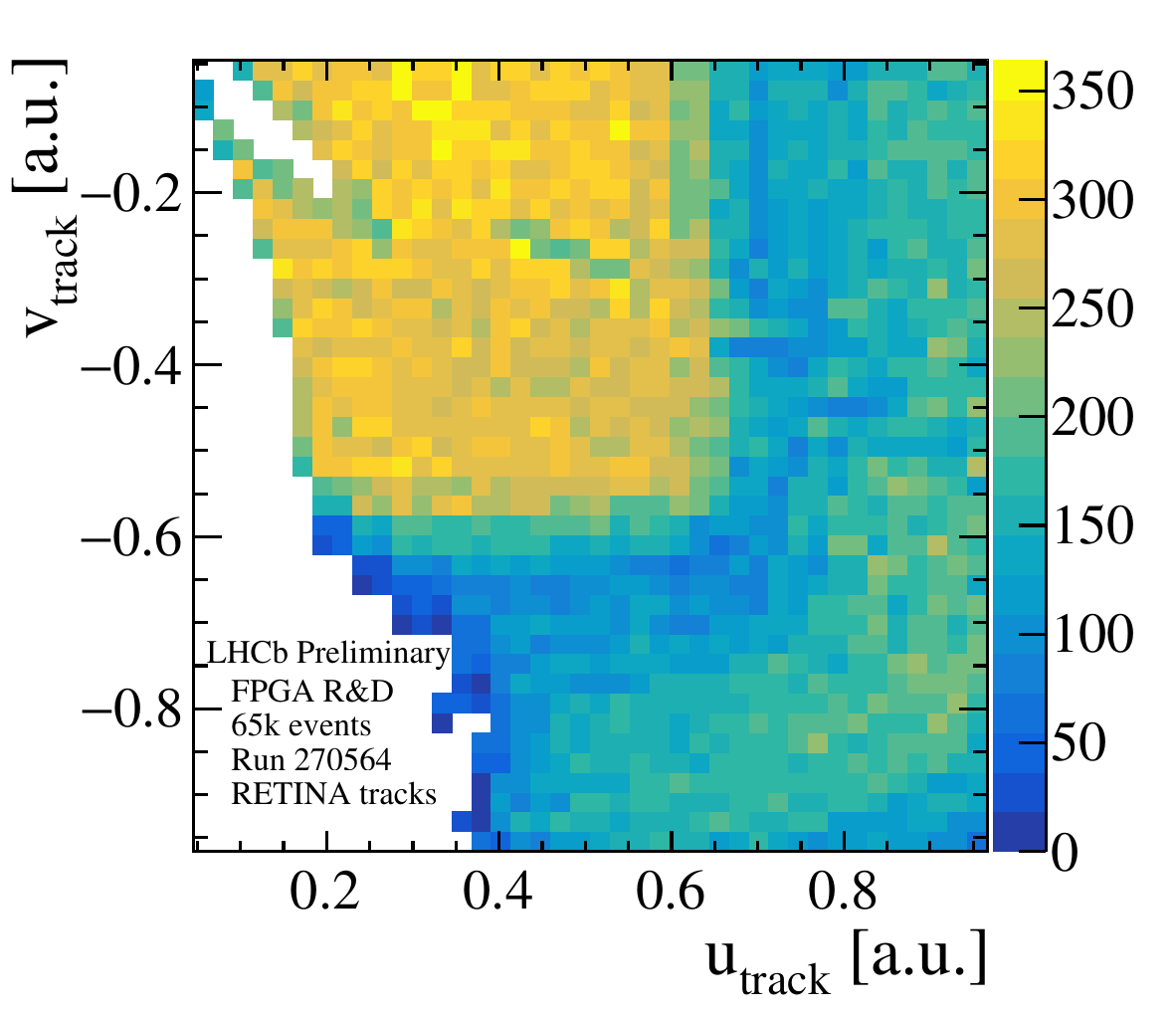}\label{subfig:demo_out}}\qquad
  \subfloat[]{\includegraphics[height=6.8cm]{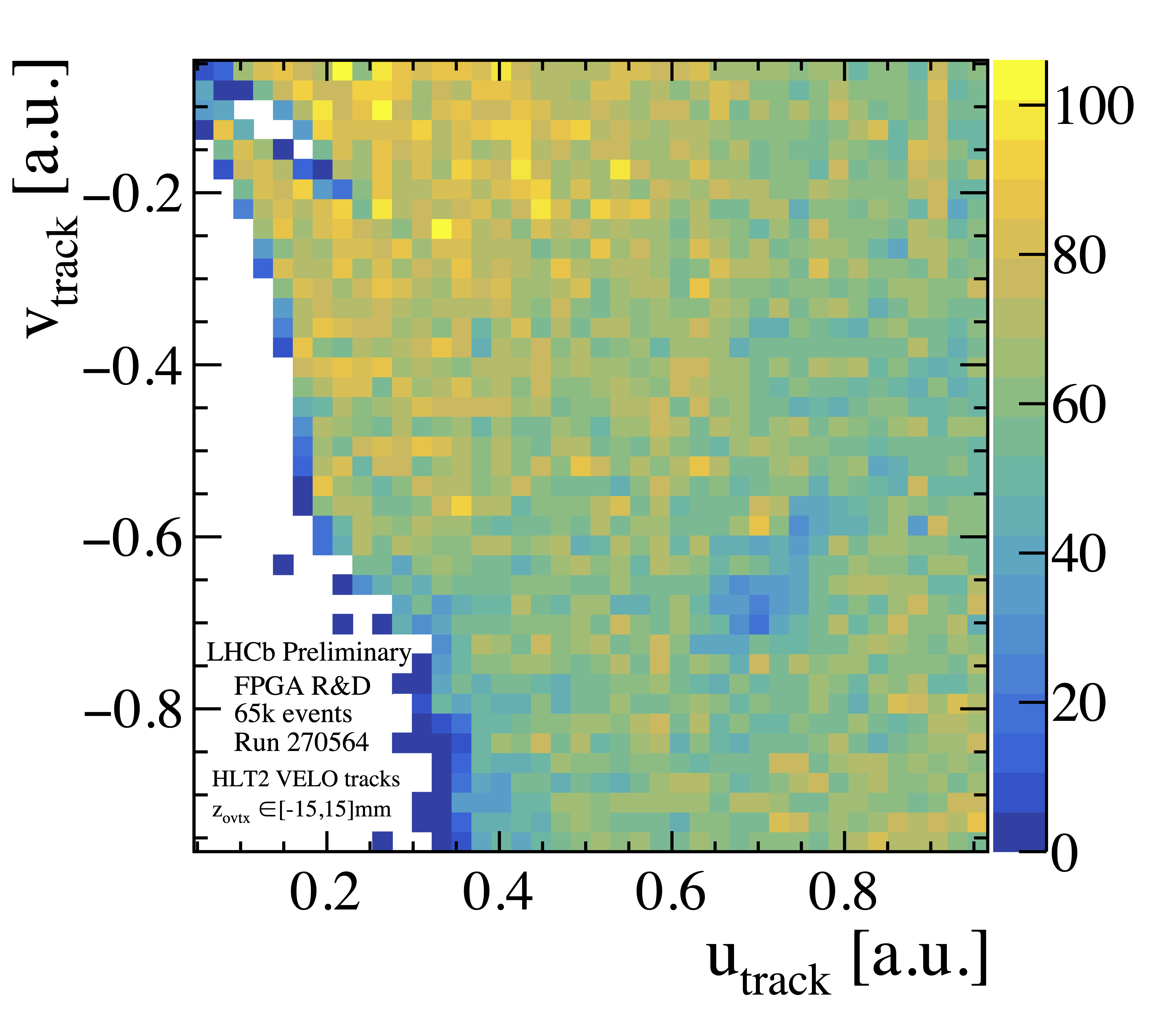}\label{subfig:hlt_out}}
  \caption{Two-dimensional distribution of the coordinates ($u-v$) of VELO tracks as reconstructed by the Retina demonstrator (a) and in the standard LHCb reconstruction inside the High Level Trigger HLT2 (b). Tracks reconstructed in HLT2 are selected to have the longitudinal position of their origin vertex compatible with $[-15,15]\si{\mm}$.}
  \label{fig:reco_tracks}
\end{figure}

For these tests, routing and receptor LUTs were re-generated to match the VELO ``open'' position in current data taking. Incidentally, the ease of this modification has provided a good test of the flexibility of the system.
The demonstrator has been taking live data in this configuration during the LHCb physics runs performed in mid-July and September. The system ran smoothly through the whole period, and the output showed no errors or any other anomalous behaviour.
Figure~\ref{fig:reco_tracks} shows the two-dimensional distribution of VELO tracks in the parameter space ($u-v$), as reconstructed by the demonstrator (a) and by standard LHCb reconstruction (b) in a $\smallsim$2 minutes sample of data taking. While a detailed event-by-event analysis of the results has not yet been possible due to the lack of persistency of the event identification number inside each file with the current setup, the two distributions appear to be in a qualitatively good agreement, especially considering that the final selection logic is not yet implemented (no acceptance cuts or clone killing are present). 

Additional tests were periodically performed during the real data run by injection of simulated data packets inside the same real-data processing chain. This allowed monitoring the functionality of the device over time, and has always returned results perfectly consistent at bit level with the output from the Retina C++ emulator, the same mentioned in Sec.~\ref{subsec:MC} and also used for physics performance studies~\cite{chep_giulia}. This supports the assumption that the system was operating correctly on real data as well. 

\section{Conclusions}
\label{sec:conclusions}
The Retina demonstrator turned out to be capable of reconstructing a quadrant of the LHCb VELO detector. 
An unparalleled event throughput of \SI{19.0}{\MHz} with good long-term reliability is achieved when processing simulated data at Run~3 conditions ($\mathcal{L} = \SI{2e33}{\invcmsqpersec}$).
The handles for reaching the desired event rate of \SI{30.0}{\MHz} with the current hardware have already been identified and are ready to be implemented.

After verifying its correct operation, the demonstrator has proved to be able to process collisions happening live at LHCb during physics data taking, without interfering with the regular DAQ. For this purpose a custom data-chain bridging the gap between the LHCb DAQ System and the demonstrator has been developed and not once in about 60 days it has crashed. The real-time reconstructed tracks appear to be sensible when qualitatively compared to the ones reconstructed by the standard LHCb software.

It thus can be concluded that the technology is mature for real-world applications in future HEP experiments. 
A proposal to the LHC scientific committee has been submitted by the LHCb collaboration for
implementing a first physics application of the Retina architecture in the upcoming Run 4 of the LHC, that will save HLT1 computing power by moving parts of the tracking reconstruction to FPGAs operating transparently at the readout level~\cite{morello_acat2019, FTDR}. 


\Acknowledgements
We gratefully acknowledge the R\&D funding received from the italian INFN without which this project would not have been possible, and the kind support of the RTA and Online groups in LHCb, that made performing these tests possible.



\end{document}